\documentclass[sigconf]{acmart}
\usepackage{amsmath}
\usepackage{booktabs}
\usepackage{graphicx}
\usepackage{float}

\usepackage{diagbox}
\usepackage{multirow}
\usepackage{subfig}
\usepackage{fancyhdr}

\AtBeginDocument{%
  }

\copyrightyear{2023} 
\acmYear{2023} 
\setcopyright{acmcopyright}
\acmConference[WSDM '23]{Proceedings of the
Sixteenth ACM International Conference on Web Search and Data
Mining}{February 27-March 3, 2023}{Singapore, Singapore}
\acmBooktitle{Proceedings of the Sixteenth ACM International Conference on
Web Search and Data Mining (WSDM '23), February 27-March 3, 2023,
Singapore, Singapore}
\acmPrice{15.00}
\acmDOI{10.1145/3539597.3570386}
\acmISBN{978-1-4503-9407-9/23/02}

\begin{document}

\title{Knowledge Enhancement for Contrastive Multi-Behavior Recommendation}

\author{Hongrui Xuan}
\affiliation{%
  \institution{Nanjing University of Aeronautics and Astronautics}
  \city{Nanjing}
  \state{}
  \country{China}
}
\email{1692595335@nuaa.edu.cn}

\author{Yi Liu}
\affiliation{%
  \institution{Nanjing University of Aeronautics and Astronautics}
  \city{Nanjing}
  \state{}
  \country{China}
}
\email{liuyi-sx21@nuaa.edu.cn}

\author{Bohan Li}
\authornote{Corresponding author}
\affiliation{%
  \institution{Nanjing University of Aeronautics and Astronautics}
  \city{Nanjing}
  \state{}
  \country{China}
}
\email{bhli@nuaa.edu.cn}

\author{Hongzhi Yin}
\affiliation{%
  \institution{The University of Queensland}
  \city{Brisbane}
  \state{}
  \country{Australia}
}
\email{h.yin1@uq.edu.au}

\renewcommand{\shortauthors}{Hongrui Xuan, Yi Liu, Bohan Li, \& Hongzhi Yin}

\begin{abstract}
  A well-designed recommender system can accurately capture the attributes of users and items, reflecting the unique preferences of individuals.
  Traditional recommendation techniques usually focus on modeling the singular type of behaviors between users and items. 
  However, in many practical recommendation scenarios (e.g., social media, e-commerce), there exist multi-typed interactive behaviors in user-item relationships, such as click, tag-as-favorite, and purchase in online shopping platforms. 
  Thus, how to make full use of multi-behavior information for recommendation is of great importance to the existing system, which presents challenges in two aspects that need to be explored: 
  (1) Utilizing users' personalized preferences to capture multi-behavioral dependencies; (2) Dealing with the insufficient recommendation caused by sparse supervision signal for target behavior. 
  In this work, we propose a Knowledge Enhancement Multi-Behavior Contrastive Learning Recommendation (KMCLR) framework, including two Contrastive Learning tasks and three functional modules to tackle the above challenges, respectively. 
  In particular, we design the multi-behavior learning module to extract users' personalized behavior information for user-embedding enhancement, and utilize knowledge graph in the knowledge enhancement module to derive more robust knowledge-aware representations for items. 
  In addition, in the optimization stage, we model the coarse-grained commonalities and the fine-grained differences between multi-behavior of users to further improve the recommendation effect. 
  Extensive experiments and ablation tests on the three real-world datasets indicate our KMCLR outperforms various state-of-the-art recommendation methods and verify the effectiveness of our method.
\end{abstract}



\maketitle

\section{Introduction}
Recommender systems aim to provide items for users personalized based on their preferences, which have emerged as critical components to alleviate information overloading in various online applications \cite{1,2,3,4}. 
How to accurately capture users' preferences from their behaviors is the key to personalized recommendations. Collaborative Filtering (CF) techniques are traditional and effective, which forecast user preference from observed behavior data.
The core idea of CF paradigm is to factorize user-item interactions into latent representations and excavate the similarity preference of individual users based on vectorized user/item representations to make recommendations \cite{5}.

With the recent boosting of deep learning, neural networks have been utilized to augment collaborative filtering architecture. 
For instance, in 2017, He et al. and Xue et al. introduced the Multi-layer Perceptron and proposed NCF \cite{5} and DMF \cite{6}, enabling the CF could handle the non-linear interaction. 
In 2018, by stacking multiple autoencoder networks \cite{7}, Du et al. designed a method to realize the high-dimensional sparse user behavior data mapping into the low-dimensional dense representations \cite{8}.
Inspired by the recent success of GNN, other research works model user-item interactions as graphs to exploit the high-order user-item relations. 
These models perform the message passing and neighborhood-based feature aggregations over the interaction graph to generate node-level embeddings (e.g., NGCF \cite{9} and GCN \cite{10}).

Nevertheless, the majority of existing methods only consider there is a singular type of behavior between users and items, which makes them unable to distill useful information from complex multiple collaborative relationships to provide sufficient recommendations. 
Actually, in many real-world scenarios, recommender systems need to deal with more complex user-item multiple interactions to comprehensively reflect the multi-dimensional users' preferences. 
For example, users can interact with items in different ways (e.g., click, tag-as-favorite and purchase) on e-commerce platforms. These behaviors characterize users' preferences from different intent dimensions, which can sever as auxiliary knowledge to provide information augmentation and help complete the recommendation tasks of the target behavior.
So how to accurately capture the dependencies between behavioral diversity and distinguish the differences between users is a challenging but valuable work. 
To tackle this challenge, many existing works have explored multi-behavior recommendation. 
For instance, ATRANK \cite{12} focuses on the feature interactions between different user behaviors in sequence recommendation, and attaches importance to a single sequence view of user's historical behavior. 
By utilizing self-attention mechanism, MATN \cite{11} realizes the encoding of pairwise correlations between different types of behaviors to recommend target behaviors.

Although the above methods are indeed effective, there are still two limitations in practical research: 
(1) target behavior data sparsity: Most recommender systems utilize supervised information for training, so it is pivotal to have enough target behavior data for the accuracy of prediction results. 
Unfortunately, in real-world scenarios, interactions observed under target behavior are often sparse compared to other types of interactions. 
It is obvious that the data of people purchasing on e-commerce platforms is much lower than the data of clicks, which leads to the direct integration of certain types of behavior that will sacrifice recommendation performance. 
(2) individualized behaviors diversity: The dependencies between different types of interactions are complex, while the interdependencies between multiple types of behaviors vary from user to user, thus leading to two problems. 
One is how to model the coarse-grained commonalities between different behaviors and another is how to model the fine-grained differences between multi-behavior of users. 
In practical scenarios, individuals exhibit different behavior patterns according to their own habits. For example, some users will frequently perform tag-as-favorite but only make sporadic purchases, while others may be not used to the operation of add-to-cart.
So how building an efficient information transfer framework to extract sufficient multi-behavioral patterns is crucial.

Recently, many studies have introduced contrastive learning (CL) into recommendation and achieved certain results \cite{16}. 
We find that CL is well suited for modeling multi-behavior information to explore coarse-grained commonalities and fine-grained differences between user-item relationships. 
Meanwhile, knowledge graphs (KG) have been widely used in recommendation and become a popular method for information completion. 
Hence, inspired by work related to CL and KG, we propose a Knowledge Enhancement Multi-Behavior Contrastive Learning Recommendation (KMCLR) framework to tackle above limitations. 
Specifically, KMCLR consists of three modules: multi-behavior learning module, knowledge enhancement module and joint learning module, which are utilized to learn user-item representations from multiple perspectives. 
Our multi-behavior learning module captures cross-type behavior dependencies, utilizing additional supervision signals from different behaviors to complement and enhance user-level information. 
While our knowledge enhancement module proposes a relation-aware knowledge aggregation mechanism, which distills additional KG information to guide the data expansion of cross-view signals and achieves item-level information enhancement. 
Finally, two different CL tasks and a loss paradigm are proposed to maximize the differences between representations of users while minimizing the gaps between different behaviors of the same user. 

In summary, the contributions of our work are highlighted as follows:

$\bullet$ We propose a new framework KMCLR for recommendation, which enhances the information from the user level and item level by combining multi-behavior messages with KG messages, emphasizing the modeling of user-item information from different perspectives, solving the issue of target behavior data sparsity.

$\bullet$ In our KMCLR framework, we propose two CL tasks and a loss paradigm, which can better model coarse-grained commonalities between behaviors and fine-grained differences between users to improve the recommendation performances. 
To the best of our knowledge, this is the first time that multi-behavior contrastive learning and knowledge graph contrastive learning are jointly introduced into recommendation.

$\bullet$ We conduct diverse experiments on three real-world datasets to show that our framework outperforms compared to various state-of-the-art recommendation methods. 
Furthermore, the ablation study is performed to better understand the model design of KMCLR and justifies the effectiveness of our key components.
\section{PRELIMINARIES}
In our studied task, we let $\mathcal{U}$ and $\mathcal{V}$ represent the set of users and items: 
$\mathcal{U}$ = $\left\{u_1,...,u_i,...,u_I\right\}$ and $\mathcal{V}$ = $\left\{e_1,...,e_j,...,e_J\right\}$, 
where I and J represent the number of users and items, respectively.
We give detailed definitions of the key notions in our KMCLR as follows:

{
\setlength{\parindent}{0cm}
\pmb{User-Item Multi-Behavior Interaction Graph.}
Since there are various types of behavior in user-item interactions, a multi-behavior interaction graph is defined as: $G_u$ = ($\mathcal{U}$, $\Theta$, $\mathcal{V}$),
where $\Theta$ = \{$\theta^1$,...,$\theta^k$ \\,...,$\theta^K$\} is the set of edge representing K types of behavior. 
Specifically, if user $u_i$ and item $v_j$ have an interaction under the type of behavior k, then $\theta^k_{i,j}$ = 1, and $\theta^k_{i,j}$ = 0 otherwise.
}

{
\setlength{\parindent}{0cm}
\pmb{Item-Item Relation Knowledge Graph.}
Considering the richness of external knowledge of items, we utilize the traditional triple method to define Knowledge Graph: $G_v = (H, R, T)$,
where $H$ denotes head-entity set, $R$ is relation set and $T$ represents tail-entity set. 
Among them, the combination of elements with the same index of these three sets forms a triple $(h, r, t)$, 
which reflects the semantic relationship $r$ between the head and tail entity $h$ and $t$.
In particular, we refer to the form of work \cite{25} to build $G_v$.
}

{
\setlength{\parindent}{0cm}
\pmb{Task Formulation.}
Based on the above definitions, we formulate the studied task as:}
\par
\pmb{Input:} 
the user-item multi-behavior interaction graph $G_u$ = ($\mathcal{U}$, $\Theta$, $\mathcal{V}$) and the item-item relation knowledge graph $G_v = (H, R, T)$.
\par
\pmb{Output:}
a prediction function which infers the interaction probability $y_{i,j}^k$ of user $u_i$ and item $v_j$ under the k-type behavior.

\begin{figure*}[ht] 
  \centering 
  \includegraphics[height=7.5cm,width=13.0cm]{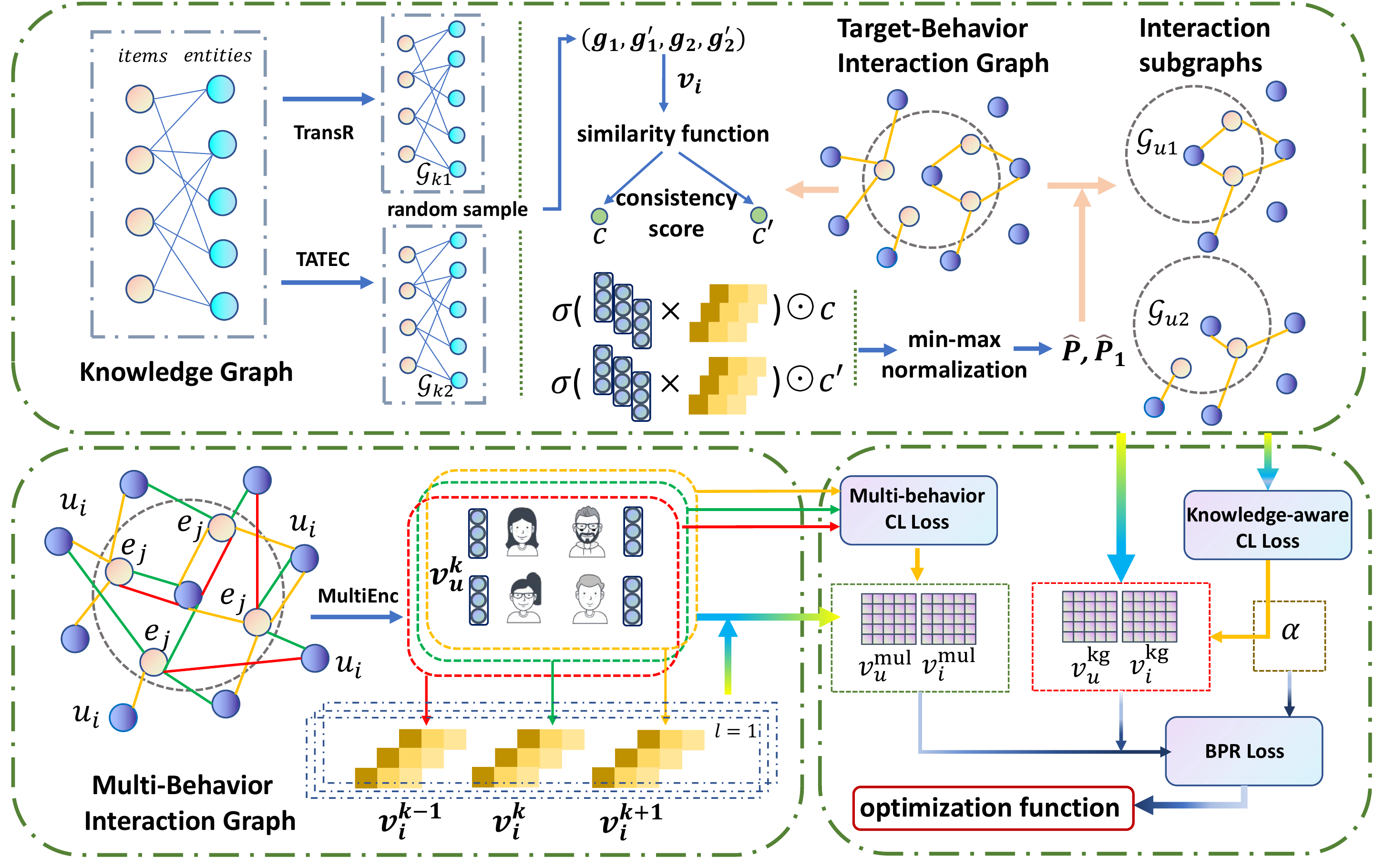} 
  \caption{The model architecture of KMCLR framework.}
  \label{Fig.1} 
\end{figure*}
\section{METHODOLOGY}
The overall architecture of the KMCLR model is shown in Figure \ref{Fig.1}. Our model mainly consists of three parts: multi-behavior learning module, knowledge enhancement module, and joint learning module. 
Two types of CL tasks are proposed to capture the multi-behavior interaction feature and item-layer knowledge feature. 
\subsection{Multi-behavior Learning Module}
In this module, the user's multi-behavior interaction graph is fed into multi-behavior encoder. we construct user single-behavior representations through this encoder. KMCLR optimizes the user-item embeddings according to the contrastive loss of other behaviors with the target behavior. This process leverages multi-behavioral information to enhance user-level embedding representations. 
\subsubsection{\textbf{Multi-behavior Information Encoding and Aggregation}}
Graph-based recommendation is usually performed on the user-item graph constructed by all behaviors, 
which can consider interactions among global multi-hops. 
To take full advantage of multi-behavior interaction graphs while injecting higher-order connectivity into multi-way relations, 
we develop a behavioral context-aware information encoder for learning single-behavior user representations:
\begin{equation}\label{eq1}
   v^k_u, v^k_i = MultiEnc^k(G_u, u, k),
\end{equation} 
where $v^k_u$ and $v^k_i$ indicate the obtained representations of user $u$ and item $i$ under the behavior $k$. 
$MultiEnc^k(\sim)$ is an optional multi-behavior encoder.
Inspired by the lightweight architecture of LightGCN \cite{15} and advanced information propagation framework MB-GMN \cite{14}, 
we utilize the following formula as $MultiEnc^k(\sim)$ to calculate the node (consist of user and item) representations:
\begin{equation}\label{eq2}
    v^{k,(l+1)}_u = \sum\limits_{i\in\mathcal{N}_u^k}v_i^{k,l}; 
    \qquad v^{k,(l+1)}_i = \sum\limits_{u\in\mathcal{N}_i^k}v_u^{k,l},
\end{equation} 
where $\mathcal{N}_u^k$, $\mathcal{N}_i^k$ are the neighboring nodes of user $u$ and item $i$. 
(l+1) means that the encoding operation is performed on the (l+1)-th layer of the neural network. Especially, we utilize the raw user and item embedding to define $v_u^{k,0}$ and $v_i^{k,0}$.
Since single-behavior representations may suffer from data sparsity issue, 
we propose to perform cross-type multi-behavior embedding aggregation, 
applying multiple layers of transformation matrices while preserving contextual information. 
The specific formulas are as follows:
\begin{equation}\label{eq3}
  v_u = PReLU((v^0_u||,...,||v^l_u||,...,||v^L_u)\times W^l),
\end{equation} 
\begin{equation}\label{eq4}
  v_u^l = \sigma(W^u \times mean(v^{1,l}_u\oplus,...,\oplus v^{k,l}_u\oplus,...,\oplus v^{K,l}_u)).
\end{equation} 
$W^u \in \mathcal{R}^{d\times d} $ and $W^l \in \mathcal{R}^{L\cdot d\times d} $ are trainable transformation parameters,
with d as latent dimension and L as the number of neural network layers.
$\sigma$ is defined as sigmoid activation function.
Similar operations are applied for item aggregation.
\subsubsection{\textbf{Multi-behavior Contrastive Learning}}
Influenced by the contrastive learning strategy, 
we hope to utilize multiple views to construct different representations and regard them as auxiliary signals to alleviate the target-behavior data sparsity. 
Meanwhile, we try to minimize the difference between multi behaviors of the same user and maximize the gap between each user through the contrastive strategy.
Therefore, a naive idea is to treat different behaviors as different views. 
To be precise, we perform contrastive learning on user embeddings between behaviors, 
and realize cross-type behavioral dependency encoding through different embedding combinations, 
completing user-level information enhancement.

One of the keys to CL is to extract valid positive-negative pair samples from different views. 
In our KMCLR, we randomly select different behaviors of user x as positive pairs, while views between users x and y are sampled as negative pairs. 
Therefore, according to the user embedding generated before, we further get the positive and negative sample pairs: 
$ \left\{v_x^k, v_x^{k'}\right\}$ and $ \left\{v_x^k, v_y^{k'}\right\}$,
where $x, y \in \mathcal{U}$ and $x \neq y$.
According to work \cite{17}, we define a self-supervised contrastive loss based on InfoNCE loss as follows:
\begin{equation}\label{eq5}
 \mathcal{L}_{MulCL} = \sum\limits_{k'=1}\limits^K\sum\limits_{x\in\mathcal{U}}-log\frac{exp(s(v_x^k,v_x^{k'})/\tau)}{\sum_{y\in\mathcal{U}}exp(s(v_x^k,v_y^{k'})/\tau)}.
\end{equation} 
$\tau$ is the parameter to control the smoothness of the softmax curve.
$s(\sim)$ denotes the pair-wise distance function to reflect the similarity of pos-neg pairs.
Through the multi-behavior CL task, KMCLR can learn more robust user representations, 
capture the underlying dependencies between target behavior and other behaviors, 
and distinguish personal preferences among different users.
\subsection{Knowledge Enhancement Module}
The knowledge graph is input into the knowledge enhancement module as auxiliary information. KMCLR utilizes different graph embedding techniques, focusing on the rationality and semantic information in the KG to obtain different node representations. Based on each graph embedding representation, an item score evaluation method from the user's perspective is proposed to construct user-item embeddings by preserving low-noise item information. These processed data will be further utilized to optimize item-level embedding representations according to CL task. 
\subsubsection{\textbf{Item-side Information Encoding and Aggregation}}
As the supplement to item-side information, 
KG contains rich relational representations on its edges. 
Under most circumstances, the importance of neighbor nodes in the KG is different. 
To reflect this difference, inspired by the graph attention mechanism \cite{18}, 
we design a method of aggregating neighbor nodes by attention weight and obtain the representations of items:
\begin{equation}\label{eq6}
  \begin{aligned} 
  &\qquad\qquad \mathcal{W}_{i,r,e} = \frac{f_{att}(v_i, v_r, v_e)}{\sum_{e'\in\mathcal{N}_i}f_{att}(v_i, v_r, v_{e'})},\\ 
  &\qquad f_{att}(v_i, v_r, v_e) = exp[\sigma(W_1(v_i||v_r||v_e))+b_1],
\end{aligned}
\end{equation} 
\begin{equation}\label{eq7}
   v_i^{l+1} = \sigma(W_2(v_i^l + \sum\limits_{e\in\mathcal{N}_i}\mathcal{W}_{i,r,e}v_e)+b_2),
\end{equation} 
where $\mathcal{W}_{i,r,e}$ denotes the weight of item $i$ to the neighbor node $e\in\mathcal{N}_i$ under the relation $r$.
Both $W_1, W_2 \in \mathcal{R}^{3d}$ and $b_1, b_2 \in\mathcal{R}$ are trainable parameters.
$v_i^{l+1}$ is the embedding of item $i$ after (l+1)-th aggregations.
Here we use $v_i^0$ to denote the raw item embedding.
\subsubsection{\textbf{Knowledge-based Enhancement and Augmentation}}
As mentioned before: we hope to construct multiple appropriate views for contrastive learning. 
Therefore, two different approaches are proposed to generate knowledge representations, 
which are the translational distance model TransR \cite{19} and the semantic matching model TATEC \cite{20}. 
Respectively, the two representations focus on the rationality and potential semantics of the triples in KG, 
so as to obtain the knowledge augmented graphs with different focus points.
The specific formulas are as follows:
\begin{equation}\label{eq8}
   f_{td}(h,r,t) = -||M_rv_h + v_r - M_rv_t||_2^2,
\end{equation} 
\begin{equation}\label{eq9}
   f_{sm}(h,r,t) = v_h^TM_rv_t + v_h^Tv_r + v_t^Tv_r +  v_h^TdDv_t.
\end{equation} 
$M_r \in\mathcal{R}^{d\times d}$ is a projection matrix from the entity space to the relation space of $r$,
while D is a diagonal matrix shared across all different relations.
$v_h, v_r, v_t$ stand for triple embeddings in KG.
So we initialize to get two different representations of the graph:
$\mathcal{G}_{k1}$ and $\mathcal{G}_{k2}$.
Then two different loss functions are exploited to optimize the node representations of KG:
\begin{equation}\label{eq10}
\mathcal{L}_{td} = \sum\limits_{(h,r,t,t')\in\mathcal{G}_{k1}}-ln\sigma(f_{td}(h,r,t')-f_{td}(h,r,t)),
\end{equation} 
\begin{equation}\label{eq11}
\qquad\mathcal{L}_{sm} = \sum\limits_{(h,r,t,t')\in\mathcal{G}_{k2}}-ln\sigma(f_{sm}(h,r,t')-f_{sm}(h,r,t)),
\end{equation} 
where $(h,r,t')$ is the triple that does not exist in KG.
Through the above methods, 
KMCLR further realizes the enhancement of knowledge representation from different perspectives

Although the introduction of KG can complement the recommendation information, 
it also introduces some noise.
Therefore, inspired by work \cite{21,22,23},
if the node representation obtained by different knowledge subgraphs is similar, 
it denotes that the node is not sensitive to structural changes with strong noise immunity.
We propose a simple method to compute graph structure consistency, reducing the effect of noisy information:
\begin{equation}\label{eq12}
    c = s(g_{1}(v_{i}), g'_{1}(v_{i})); \qquad(g_{1}, g'_{1} \in \mathcal{G}_{k1}^{sub}),
\end{equation} 
\begin{equation}\label{eq13}
    c' = s(g_{2}(v_{i}), g'_{2}(v_{i})); \qquad(g_{2}, g'_{2} \in \mathcal{G}_{k2}^{sub}).
\end{equation} 
Here, $(g_{1}, g'_{1}, g_{2}, g'_{2})$ are subgraphs sampled from $\mathcal{G}_{k1}$ and $\mathcal{G}_{k2}$ by setting different seeds.
$s(\sim)$ is the similarity function to calculate the consistency parameter $c \in \mathcal{R}^{1\times M}$ for all item nodes.
Specifically, we obtain the node representations $v_i$ of subgraphs by Equation (\ref{eq7}).
Generally speaking, the stronger stability and noise resistance of a node, the higher its corresponding score $c$, 
which is used as guiding information to optimize the user-item interaction graph. 

Stable nodes benefit to resist noise, but not enough to reflect user preferences. 
Some users may prefer to interact with nodes of low stability. 
So, after obtaining the consistency score of nodes in KG, 
we attempt to adjust the score from user's perspective and utilize the score to guide the selection of edges in interaction graph. 
We designed the following sampling probability calculation formulas:
\begin{equation}\label{eq14}
    P_{u,i} = \sigma(v_u^Tv_i)\odot c,
\end{equation} 
\begin{equation}\label{eq15}
    \hat{P_{u,i}} = (1-Min\_Max(P_{u,i}))a + Min\_Max(P_{u,i})b.
\end{equation} 
$P_{u,i}$ is the initial sampling probability based on user's interactions and consistency score $c$, 
while $\odot$ is the element-wise multiplication with a broadcast mechanism.
KMCLR performs min-max normalization function $Min\_Max(\sim)$ to transform $P_{u,i}$ into interval [a,b] to get final sampling probability of interaction graph.
According to this probability, 
the edge retention in the interaction graph is determined, 
obtaining two user-item subgraphs $\mathcal{G}_{u1}$ and $\mathcal{G}_{u2}$.

\subsubsection{\textbf{Knowledge-aware Contrastive Learning}}
After the above steps, KMCLR obtains two pairs of perspective graphs with different emphasis information:
$(\mathcal{G}_{u1},\mathcal{G}_{k1})$ and $(\mathcal{G}_{u2},\mathcal{G}_{k2})$.
According to the Equation (\ref{eq7}), the representations of items are encoded to preserve the information of KG.
Then, similar to the multi-behavior learning module, we utilize LightGCN to perform message propagation and calculate the embeddings of user-item graph.
The specific formula is the same as Equation (\ref{eq2}).

Based on the two pairs of perspectives mentioned above, 
KMCLR performs contrastive learning on view-specific node representations, 
which treats the same node embeddings from different views as positive pairs and different nodes from different views as negative pairs.
Similar to Equation (\ref{eq5}), we define the following formula:
\begin{equation}\label{eq16}
     \mathcal{L}_{KCL} = \sum\limits_{x\in\mathcal{U\cup V}}-log\frac{exp(s(v_x^1,v_x^2)/\tau)}{\sum_{y\in\mathcal{U\cup V}}exp(s(v_x^1,v_y^{2})/\tau)}.
\end{equation} 

Through the knowledge-aware CL task, 
KMCLR can supplement the item-side information from the external knowledge graph, 
reduce the impact of external noise, 
and alleviate the issue of data sparsity to a certain extent.

\subsection{Joint Learning Module}
Through the above modules, we obtained two types of user-item embeddings generated in different ways, 
one using multi-behavior information and the other using KG information.
These embeddings obtained from different perspectives contain messages with different emphases, 
which require to be combined in a reasonable way.
Therefore, we assign a combination weight $\alpha$ and explore the optimal combination result through experiments.
We can get:
\begin{equation}\label{eq17}
     v_u = (1-\alpha)v_u^{mul} + \alpha v_u^{kg},\qquad v_i = (1-\alpha)v_i^{mul} + \alpha v_i^{kg}.
\end{equation} 

During the optimization phase,
following the classical ranking model \cite{26}, 
KMCLR leverages the Bayesian Personalized Ranking (BPR) recommendation loss as the main task loss to further optimal parameters:
\begin{equation}\label{eq18}
     \mathcal{L}_{BPR} = -\sum\limits_{u\in \mathcal{U}}\sum\limits_{i\in \mathcal{N}_{u}}\sum\limits_{j\notin \mathcal{N}_{u}}In\sigma(v_u^Tv_{i}-v_u^Tv_{j}),
\end{equation} 
where $\mathcal{N}_u$ denotes the set of observed interactions of users $u$.

After defining all three losses, 
a training strategy paradigm is proposed to better learn information from different views and further optimize the model.
Specifically: (1) First, we will train the two contrastive learning models separately in the multi-behavior learning module and the knowledge enhancement module, 
learning different initial parameter spaces of the two modules on separate training data. 
(2) Then, the two-part trained node embeddings are concatenated with the combined weight $\alpha$. 
We calculate the main model BPR loss and CL loss for the new node embeddings as the final optimization step of KMCLR. 
It should be noted that after study, 
we decided to utilize LightGCN in the multi-behavior learning module as our main model.
The specific training paradigm is as follows:
\begin{equation}\label{eq19}
  \begin{aligned}
   & arg min\triangleq main\_opt\{mul\_opt(L^{mul},V^{mul}),\\
   & \qquad\qquad kg\_opt(L^{kg},V^{kg}), V^{main}\}.
  \end{aligned}
\end{equation} 

Here $\_opt$ is the parameter optimizer, 
$L$ and $V$ are the loss and node representation set of the corresponding module, respectively.
Among them, for the loss $L$, we give the following definition:
\begin{equation}\label{eq20}
    L = \mathcal{L}_{BPR} + \lambda1\mathcal{L}_{cl} + \lambda2||\Omega||_2^2,
\end{equation} 
where $\Omega$ is the learnable model parameters, 
while $\lambda1$ and $\lambda2$ are utilized to control the weights of self-supervised signals and regularization terms.

\subsection{Model Complexity Analysis}
We analyze the complexity of KMCLR from the following aspects:
(1) In the multi-behavior learning module, the cost of lightweight neural architecture is $O(L \times K \times |\Theta^+| \times d)$ 
with $L$ network layers, $K$ types of behavior, $|\Theta^+|$ edges set of interactions and $d$ hidden dimensionality.
The cross-type of behavior aggregation costs $O(L \times K \times (I+J) \times d)$.
The complexity to calculate InfoNCE loss is $O(B \times K \times |\Theta^+| \times d)$, where $B$ is the sample size in a epoch of training.
(2) In the knowledge enhancement module, the operations of information aggregation takes $O(|R| \times d)$ with $|R|$ as the number of relations in $G_v$.
For the generation of each pair of perspective graphs, KMCLR costs $O(|\Theta^+| \times d^2 + |\Theta^+| \times d \times L)$.
The complexity of InfoNCE-based loss calculation is $O(B \times d \times (I+J))$. 
Compared to existing models, KMCLR improves performance by slightly sacrificing complexity.

\section{EXPERIMENTS}

In this section, we evaluate the performance of KMCLR through extensive experiments and answer the following questions: 
\textbf{(RQ1)} How does KMCLR perform compared to other recommendation methods? 
\textbf{(RQ2)} What are the effects of different modules in KMCLR on performance? 
\textbf{(RQ3)} How does KMCLR effectively alleviate data sparsity issue for recommendation? 
\textbf{(RQ4)} How different hyperparameter settings affect the performance? 
\textbf{(RQ5)} How does KMCLR alleviate the noise issue caused by extra information?

\subsection{Datasets for Experiments}
We validate the effectiveness of KMCLR with experiments on three real-world datasets.
\textbf{Yelp:} This dataset is collected by well-known business review websites in the US, 
covering restaurants, shopping malls, hotels and other fields, 
usually used for business venue recommendation. 
According to the settings of \cite{25}, 
we considered four behaviors (i.e., \emph{dislike}, \emph{neutral}, \emph{tip}, \emph{like}).
\textbf{Tmall:} It is collected by one of the largest E-commerce platforms in China. 
In this dataset, 
user interactions include \emph{page view}, \emph{tag-as-favorite}, \emph{add-to-cart} and \emph{purchase}.
\textbf{Retail:}  It is a dataset about online retail scenarios with explicit multi-type user-item interactions. 
We extracted the following three behaviors as training data, 
i.e., \emph{tag-as-favorite}, \emph{add-to-cart} and \emph{purchase}.
Table \ref{tb1} presents the detailed statistics of the above datasets.
Furthermore, referring to previous studies, we cite the knowledge graph constructed in work \cite{25} during the experiment.
\begin{table}{
  \setlength{\tabcolsep}{2pt}
  \caption{Statistics of datasets}
  \label{tb1}
  \small
  \begin{tabular}{ccccc}
    \toprule
    Dataset&Users&Items&Interactions&Interactive Behavior Type\\
    \midrule
    Yelp &19,800&22,734&1,400,002& \{Dislike, Neutral, Tip, Like\}\\
    Tmall &31,882&31,232&1.451,219& \{Page View, Favorite, Cart, Purchase\}\\
    Retail &147,894&99,037&1,584,238& \{Favorite, Cart, Purchase\}\\
  \bottomrule
  \vspace{-2.0em}
\end{tabular}
}
\end{table}
\begin{table*}[ht]
  \center
  \caption{Performance comparison on different datasets in terms of HR@10 and NDCG@10}
  \label{tb2}
  \resizebox{\linewidth}{!}{
  \begin{tabular}{|c|c|c|c|c|c|c|c|c|c|c|c|c|c|}\hline
  \multicolumn{1}{|c|}{Data}
   &Metric&BPR&LightGCN&AutoRec&NGCF&KGAT&KGCL&NMTR&MATN&MBGCN&KHGT&CML&KMCLR\\ \hline
  \multirow{2}{*}{Yelp}&HR&0.7435&0.7884&0.7592&0.7903&0.8211&0.8331&0.7880&0.8210&0.7962&0.8767&\underline{0.8774}&\textbf{0.8897}\\ \cline{2-14}
  \multirow{2}{*}{ }&NDCG&0.4497&0.4993&0.4705&0.5063&0.5362&0.5516&0.4673&0.5331&0.5033&\underline{0.5970}&0.5965&\textbf{0.6038}\\ \hline
  \multirow{2}{*}{Tmall}&HR&0.2443&0.3418&0.3274&0.3288&0.3853&0.4027&0.3623&0.4302&0.3910&0.4032&\underline{0.5185}&\textbf{0.5671}\\ \cline{2-14}
  \multirow{2}{*}{ }&NDCG&0.1501&0.2048&0.1903&0.1964&0.2219&0.2293&0.2097&0.2437&0.2261&0.2351&\underline{0.3055}&\textbf{0.3540}\\ \hline
  \multirow{2}{*}{Retail}&HR&0.2608&0.3065&0.2779&0.3027&0.3485&0.3502&0.3112&0.3490&0.3137&0.4199&\underline{0.4256}&\textbf{0.4557}\\ \cline{2-14}
  \multirow{2}{*}{ }&NDCG&0.1653&0.1847&0.1683&0.1844&0.2191&0.2197&0.1869&0.2183&0.1904&0.2471&\underline{0.2503}&\textbf{0.2735}\\ \hline
  \end{tabular}
  }
\end{table*}
\subsection{Methods for Comparison}
We compare KMCLR with various baselines from different research lines as follows:\\
\\\textbf{Collaborative Filtering-Based Method:}
\\$\bullet$ \textbf{BPR \cite{26}:} It is a traditional collaborative filtering method for personalized ranking of items with pairwise ranking loss.
\\ $\bullet$ \textbf{LightGCN \cite{15}:} It uses GNN for collaborative filtering and simplifies convolution operations between message propagation.
\\ $\bullet$ \textbf{AutoRec \cite{27}:} It is an autoencoder-based model that projects user-item interactions into a hidden low-dimensional space.
\\ $\bullet$ \textbf{NGCF \cite{28}:} It is a message passing framework based on graph neural network collaborative filtering, which utilizes high-order connectivity for aggregation.
\\
\\\textbf{Knowledge-Aware Recommendation Method:}
\\ $\bullet$ \textbf{KGAT \cite{29}:} It designs an attentive message propagation framework based on knowledge collaboration graph to study higher-order connectivity between semantic relations.
\\ $\bullet$ \textbf{KGCL \cite{22}:} It utilizes CL as auxiliary information to combine KG learning with user-item interaction modeling under the joint self-supervised learning paradigm.
\\\textbf{Multi-Behavior Recommendation Method:}
\\ $\bullet$ \textbf{NMTR \cite{30}:} It is a multi-task recommendation framework that can cascade predictions on multiple types of behaviors.
\\ $\bullet$ \textbf{MATN \cite{11}:} It employs attention networks and memory units to learn interactive dependencies between multiple types of behaviors.
\\ $\bullet$ \textbf{MBGCN \cite{13}:} It is a graph convolutional network-based model that propagates behavior-aware embeddings on a unified graph by modeling user multiple behaviors.
\\ $\bullet$ \textbf{KHGT \cite{25}:} It introduces a transformer-based approach to multi-behavior modeling with emphasis on temporal information and auxiliary knowledge information, modeling behavior embeddings through graph attention networks.
\\ $\bullet$ \textbf{CML \cite{1}:} It introduces CL into multi-behavior recommendation, proposing meta-contrastive coding to enable the model to learn personalized behavioral features.
\subsection{Parameter Settings}
We implement KMCLR with Pytorch and set the users and items embedding sizes of all modules to 32. 
Embedding initialization is performed in the knowledge enhancement module and the multi-behavior learning module via the Normal and Xavier methods, respectively. 
We optimize all models by Adma optimizer with the learning rate of $1e^{-3}$ and $3e^{-4}$ for different modules mentioned above. 
We apply L2 regularization with decaying weights in the range $\left\{0.05, 0.01, 0.005, 0.001, 0.0005, 0.0001\right\}$ to learn embeddings.
The number of message propagation layers utilized in the model is selected from $\left\{1, 2, 3, 4\right\}$.
The combination weight $\alpha$ is set to choose from the range $\left\{0.1, 0.2, 0.3, 0.4\right\}$.
For the compared baseline experiments, the parameters are set according to the original text.
\subsection{Performance Validation (RQ1)} 
We present the performance evaluation results of all baseline methods on different datasets in Table \ref{tb2}, 
and summarize the following observations:

Compared with all other baselines, KMCLR achieves the best performance on each dataset, 
which confirms the effectiveness of our model. Since various datasets have large differences in data size and sparsity, 
it proves that KMCLR has good generality. We mainly attribute the significant improvement on performance to the following two aspects: 
(1) Benefiting from the introduction of multi-behavior information and additional KG, 
the model can fully capture personalized information from multiple aspects as signal to guide the recommendation of target behavior. 
(2) Due to the introduction of auxiliary self-supervised tasks, the model can excavate individual preferences in terms of higher-level features, 
clearly distinguishing the information gradient between individuals.

We can notice that the knowledge-aware recommendation and multi-behavior recommendation are always better than the single-behavior recommendation based on collaborative filtering, 
which means that it is beneficial to introduce additional auxiliary information into the personalization modeling, 
which can be to a certain extent making up for the issue of sparse data in a single behavior. In addition, among various baselines, 
CML is usually the one with the best performance. 
This result indicates that the introduction of contrastive learning in multi-behavior recommendation can make the model better learn the commonalities and differences between individual behaviors, 
through different semantics guiding more refined recommendations. 
Furthermore, KHGT also achieves sub-optimal results in most cases, proving the effectiveness of jointly considering multi-behavior views and knowledge graph views.
\begin{table}[htbp]
  \setlength{\tabcolsep}{2pt}
    \caption{Results of ablation experiments}\label{table3}
    \centering
\begin{tabular}{c|cc|cc|cc}
  \hline
  \multirow{2}*{\diagbox{model}{dataset}} & \multicolumn{2}{c|}{Yelp} & \multicolumn{2}{c|}{Tmall} & \multicolumn{2}{c}{Retail}                                                     \\
  &HR&NDCG&HR&NDCG&HR&NDCG\\
  \hline
  w/o-Mcl&0.8352&0.5749&0.4583&0.2645&0.3612&0.2157 \\
  w/o-Kcl&0.8687&0.5916&0.5231&0.3064&0.4247&0.2513 \\
  w/o NorT&0.8701&0.5943&0.5339&0.3106&0.4268&0.2502\\
  \hline
  KMCLR&\textbf{0.8897}&\textbf{0.6038}&\textbf{0.5671}&\textbf{0.3540}&\textbf{0.4557}&\textbf{0.2735}\\
  \hline
\end{tabular}
    \vspace{-1.0em}
\end{table}
\subsection{Model Ablation Study (RQ2)} 
To explore the impact of different components on performance improvement and demonstrate its effectiveness, 
we consider three variants of KMCLR and design the following ablation experiments:
\\ $\bullet$ \textbf{w/o-Mcl:} In order to study the influence of multi-behavior information on the model, 
we propose a variant w/o-Mcl, which disables the multi-behavior learning module, 
and only utilizes the information of a single target behavior for embedding generation and learning. 
As can be seen from Table \ref{table3}, 
compared with the complete KMCLR, the performance of the version without the multi-behavior learning module is significantly reduced in both indicators. 
It is because behaviors generated by the same user can profile individual preferences in multiple ways. 
Utilizing information contrast to model coarse-grained commonalities between different behaviors helps KMCLR learn better representations while making it distinguish fine-grained differences between individuals. 
The above-mentioned aspects can also be used as supplementary messages to deal with the issue of data sparsity.
\\ $\bullet$ \textbf{w/o-Kcl:} To confirm that the introduction of KG is beneficial to enhance the item-side information and plays a positive role in target recommendation, 
we design another variant w/o-Kcl. 
This variant makes KMCLR degenerate into a simple multi-behavior contrastive learning model by disabling the knowledge enhancement module. 
The experimental results show that the introduction of KG as auxiliary information will improve the performance of the model, 
which is in line with common sense. Although the introduction of KG may additionally lead to an increase in noisy information, 
our designed knowledge enhancement module minimizes this effect by selecting stable nodes. 
Meanwhile, KMCLR processes the KG to generate two graphs with different focus to strengthen the role of knowledge on the final recommendation from different perspectives.
\\ $\bullet$ \textbf{w/o NorT:} During the training phase of the model, we propose a joint training paradigm to enhance the learning effect of KMCLR. 
To verify its effect, we adopt a normal training strategy in ablation experiments, 
i.e., directly optimizing the linear weighted sum of two contrastive losses and BPR loss. 
This normal optimization does not consider the integration method between the auxiliary task and the main task, 
ignoring the impact of different self-supervised learning on the main function. 
In contrast, KMCLR's strategy is to first train different views' node embeddings individually, 
and then ensemble optimization. The auxiliary information is integrated into the node embeddings from the bottom layer through multiple rounds of optimization. 
Experimental results show that our training strategy is more conducive to the recommendation of target behavior.
\begin{figure}[!htb]
    \vspace{-1.0em}
    \centering
    \subfloat[{\centering{Retail HR@10}}]{
      \includegraphics[height=3.2cm, width=4cm]{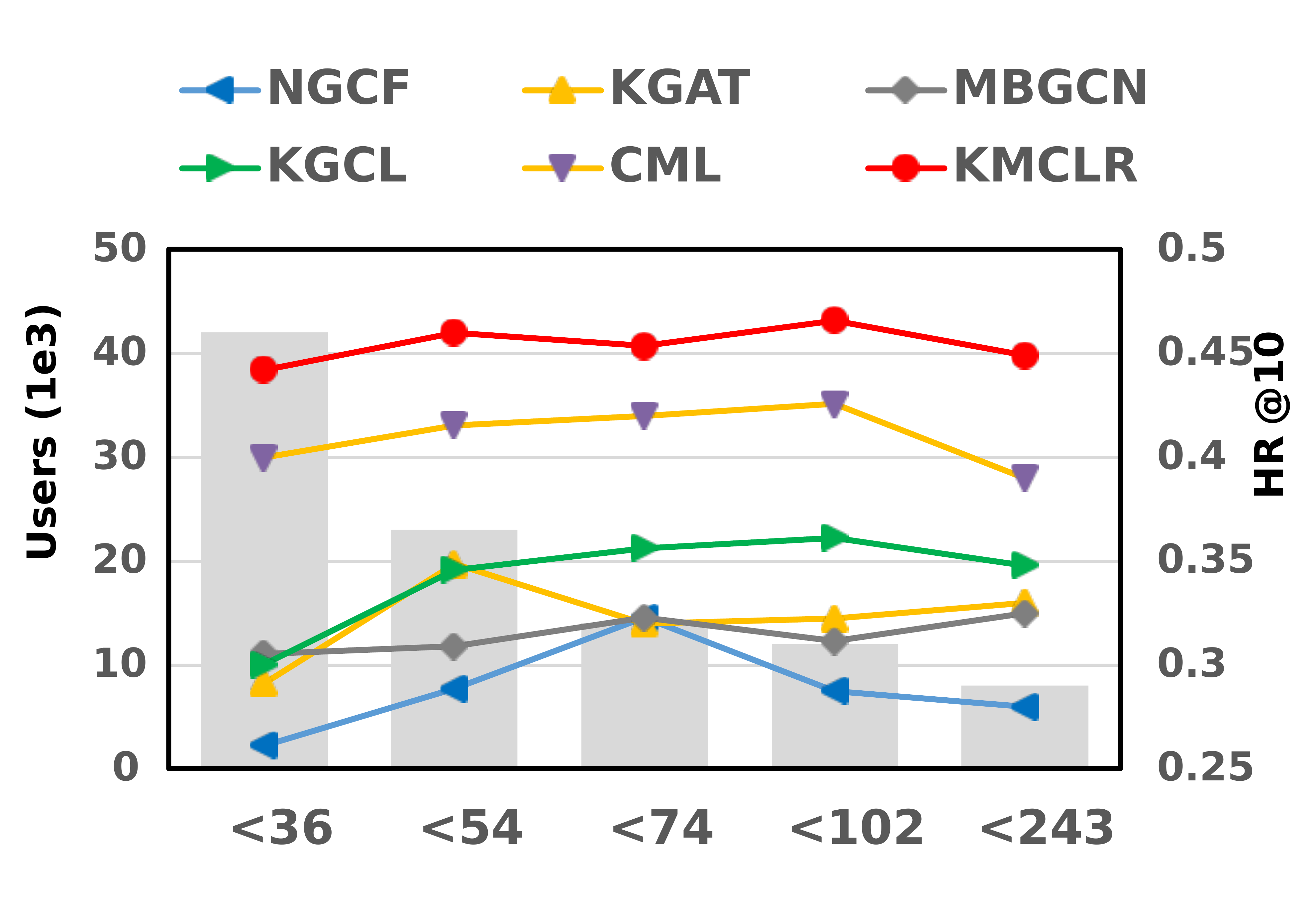}}
    \subfloat[{\centering{Retail NDCG@10}}]{
      \includegraphics[height=3.2cm, width=4cm]{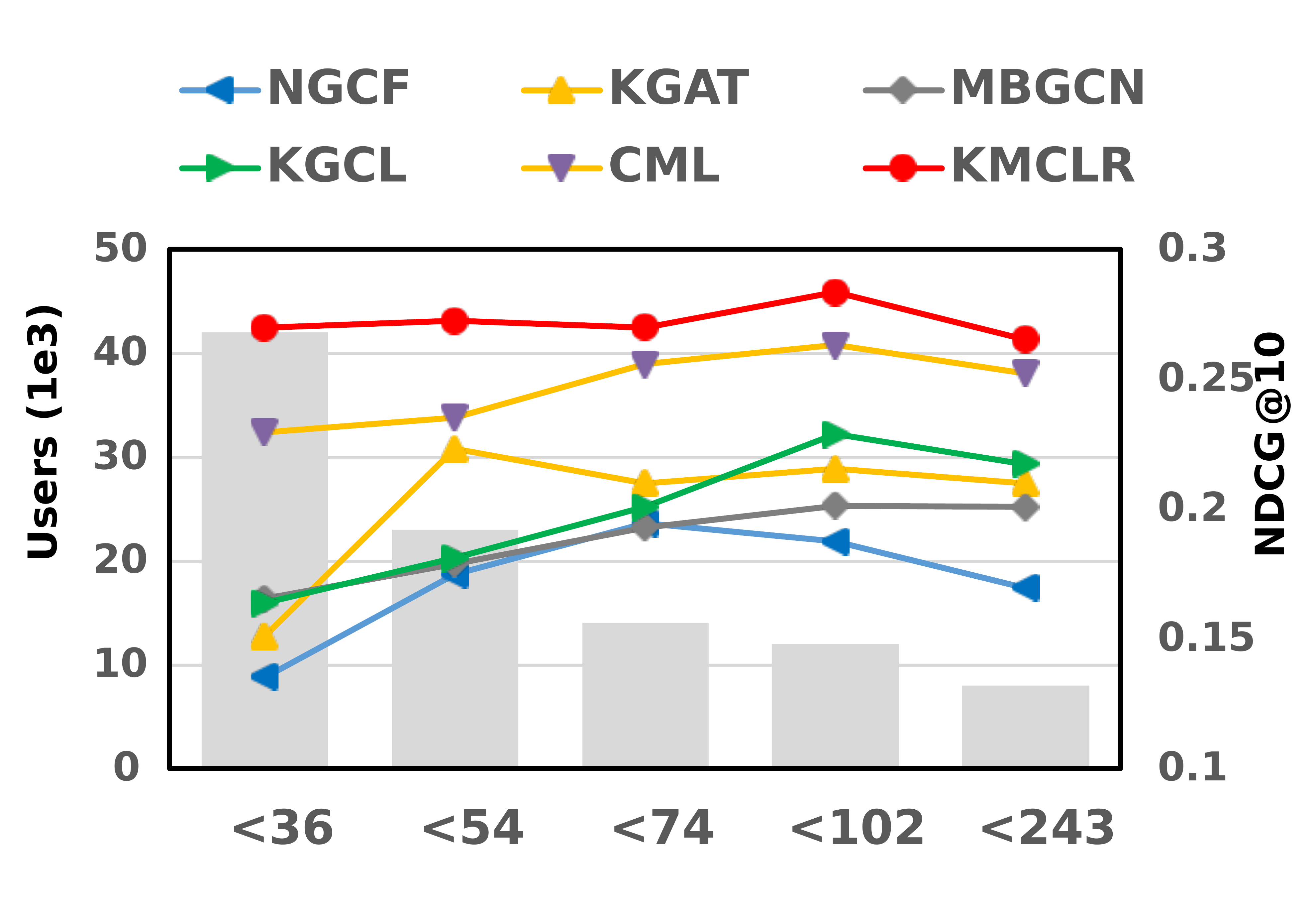}}
    \caption{\textbf{Performance of KMCLR and baseline methods w.r.t different data sparsity degrees on Retail data.}}
    \label{Data Sparsity}
    \vspace{-1.0em}
  \end{figure}
\subsection{Performance on Alleviating Data Sparsity (RQ3)} 
Since the issue of interactive data sparsity is prevalent in real-world scenarios \cite{44}, 
the ability to alleviate this issue becomes a metric for evaluating a recommender system. 
Additional experiments are conducted to verify the performance of models on interactive data with varying degrees of sparsity. 
We follow similar settings in \cite{25} to generate datasets of varying degrees of sparsity for Retail. 
As shown in Figure \ref{Data Sparsity}, 
the x-axis indicates the number of user interactions, and the y-axis represents the recommendation accuracy. 

We conclude that: (1) The collaborative filtering-based model (NGCF) is greatly affected by data sparsity, 
so combining additional messages (e.g., multi-behavior information, knowledge information) can mitigate this effect to a certain extent. 
(2) Compared with simply introducing auxiliary information (KGAT, MBGCN), the model combined with contrastive learning (KGCL, CML) is more effective, 
demonstrating that CL enables the model to maintain the heterogeneity of behavior and the diversity of knowledge semantics through the self-supervised paradigm. 
(3) Compared with other methods, KMCLR performs consistently when dealing with data with different degrees of sparseness. 
The results indicate that the performance gap between our approach and other competitors becomes more significant as data becomes sparser, 
which proves that the joint introduction of multi-behavior information and KG information can better alleviate the data sparsity, 
allowing the model to consider multi-view user preferences by incorporating contextual (e.g., click, tag-as-favorite) and auxiliary information.
\begin{figure}[!htb]
  \vspace{-1.0em}
  \centering
  \subfloat{
    \includegraphics[height=2.5cm,width=4cm]{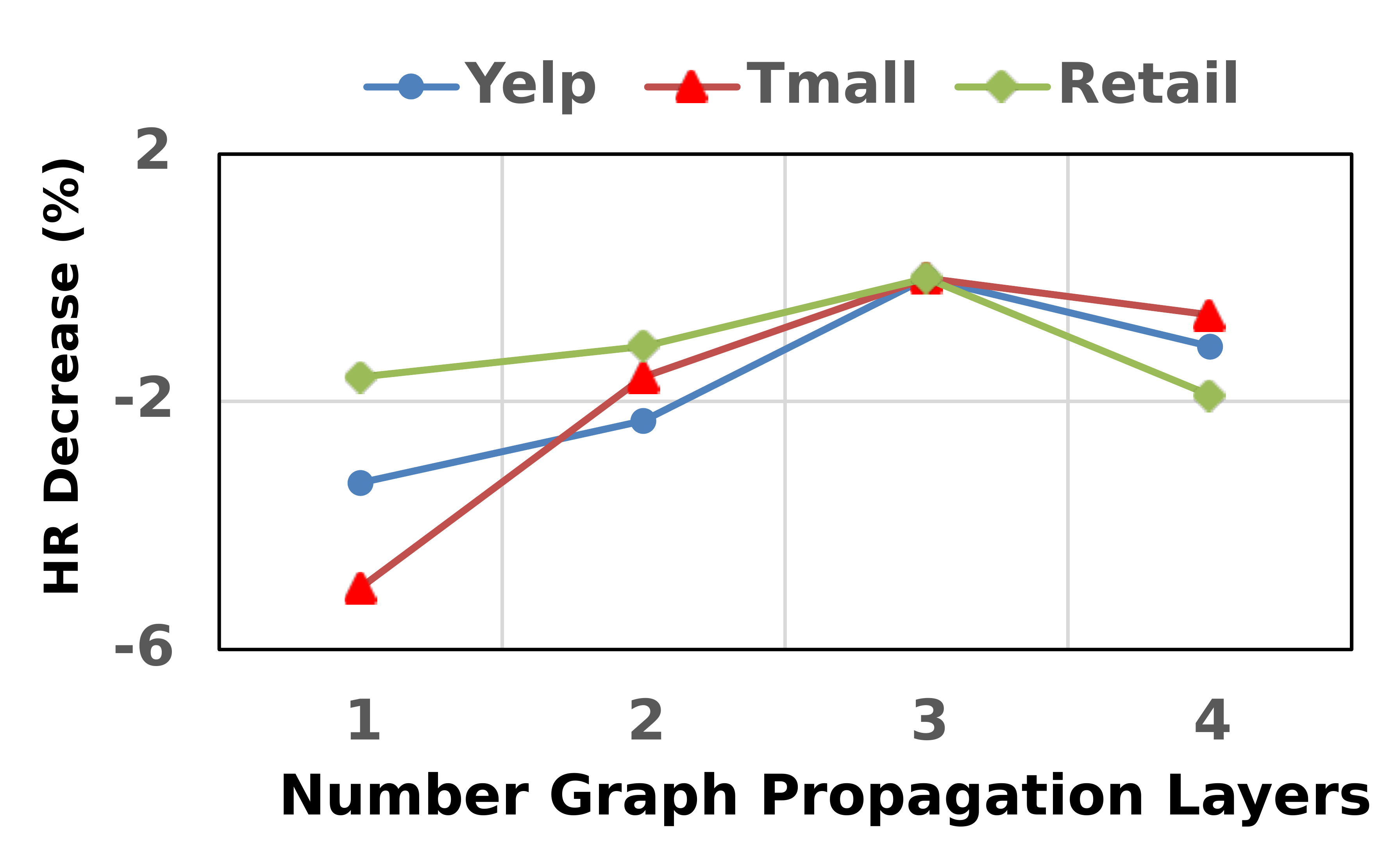}}
  \subfloat{
    \includegraphics[height=2.5cm,width=4cm]{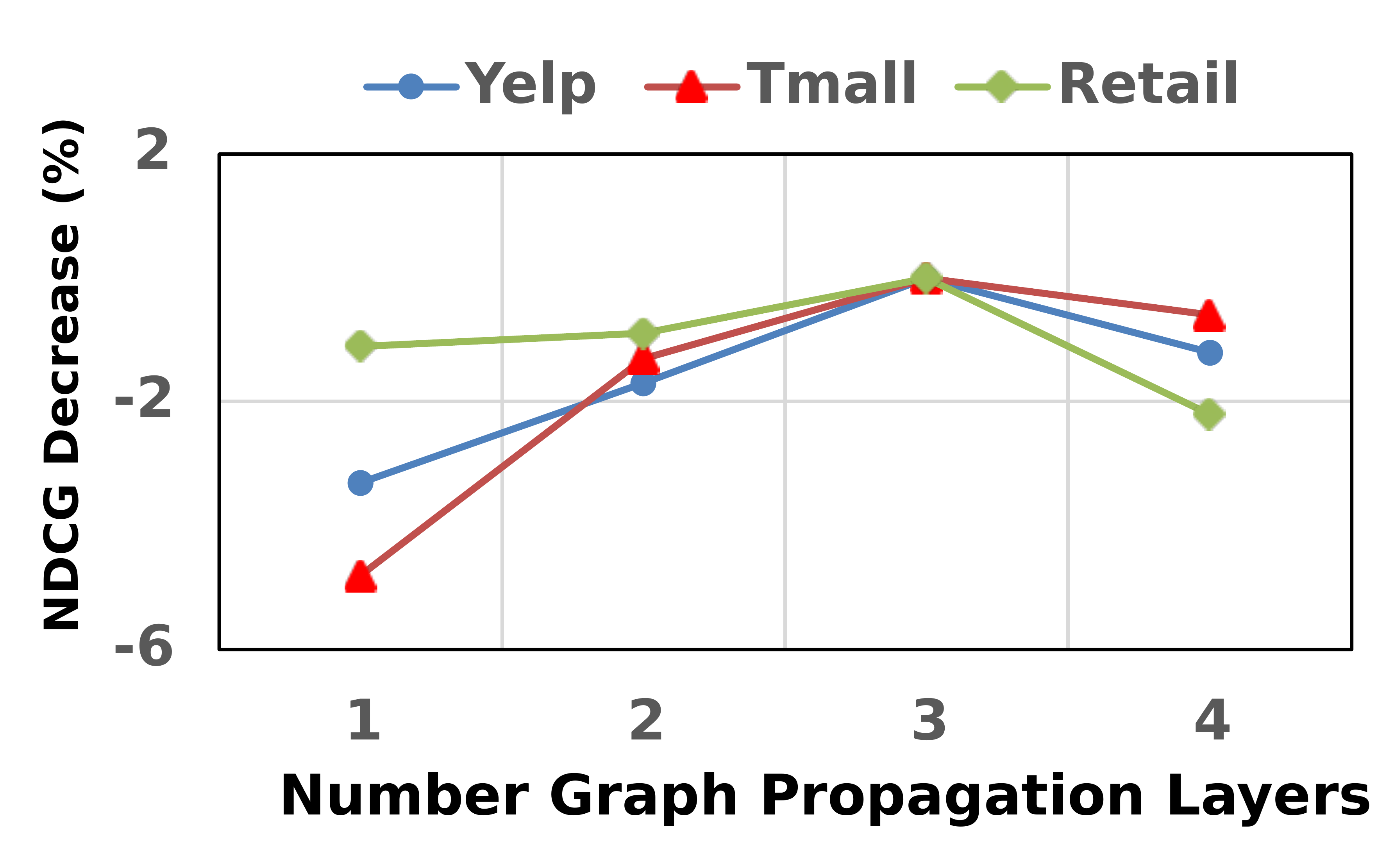}}
  \vspace{-5mm}
  
  \subfloat{
    \includegraphics[height=2.5cm,width=4cm]{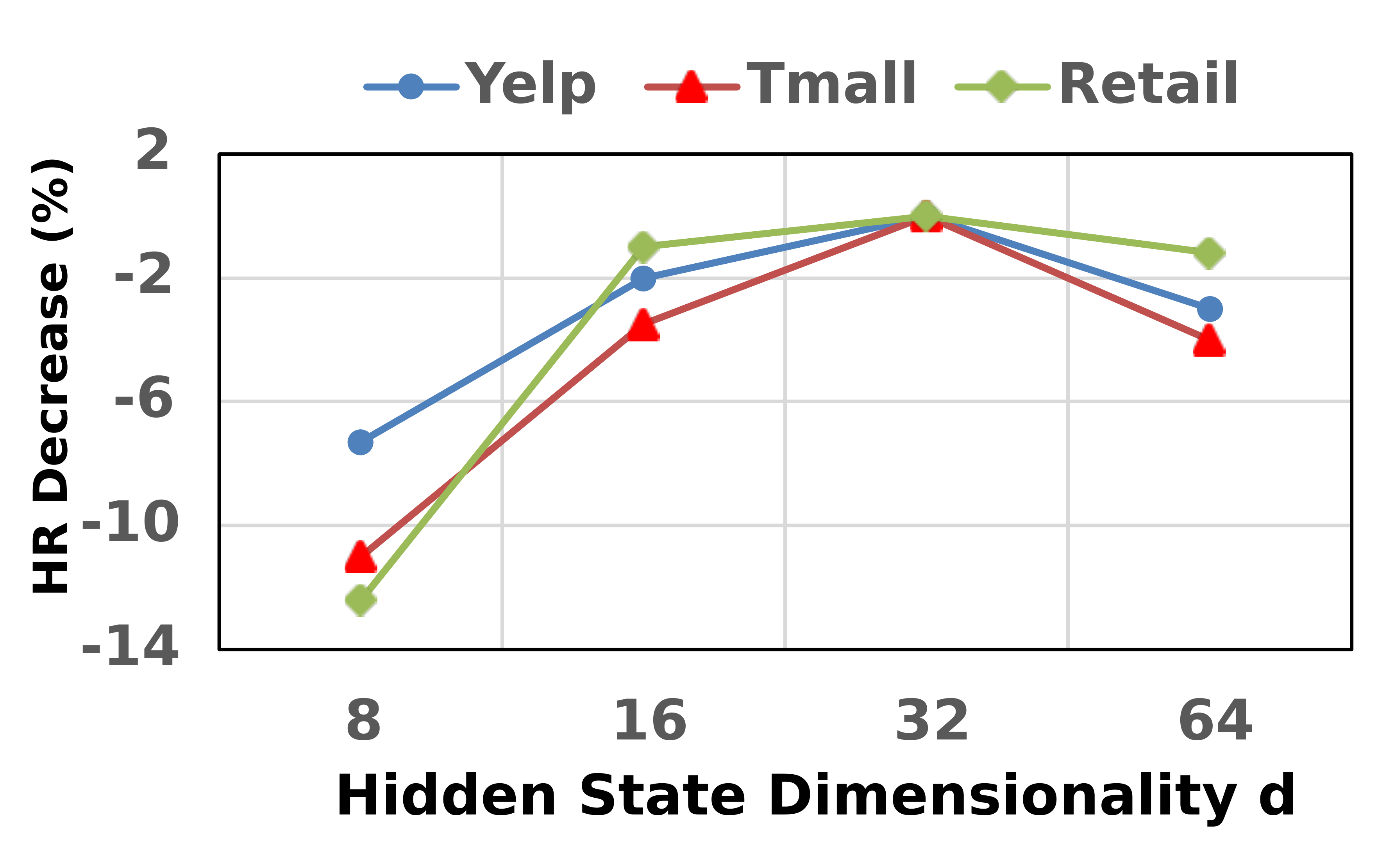}}
  \subfloat{
    \includegraphics[height=2.5cm,width=4cm]{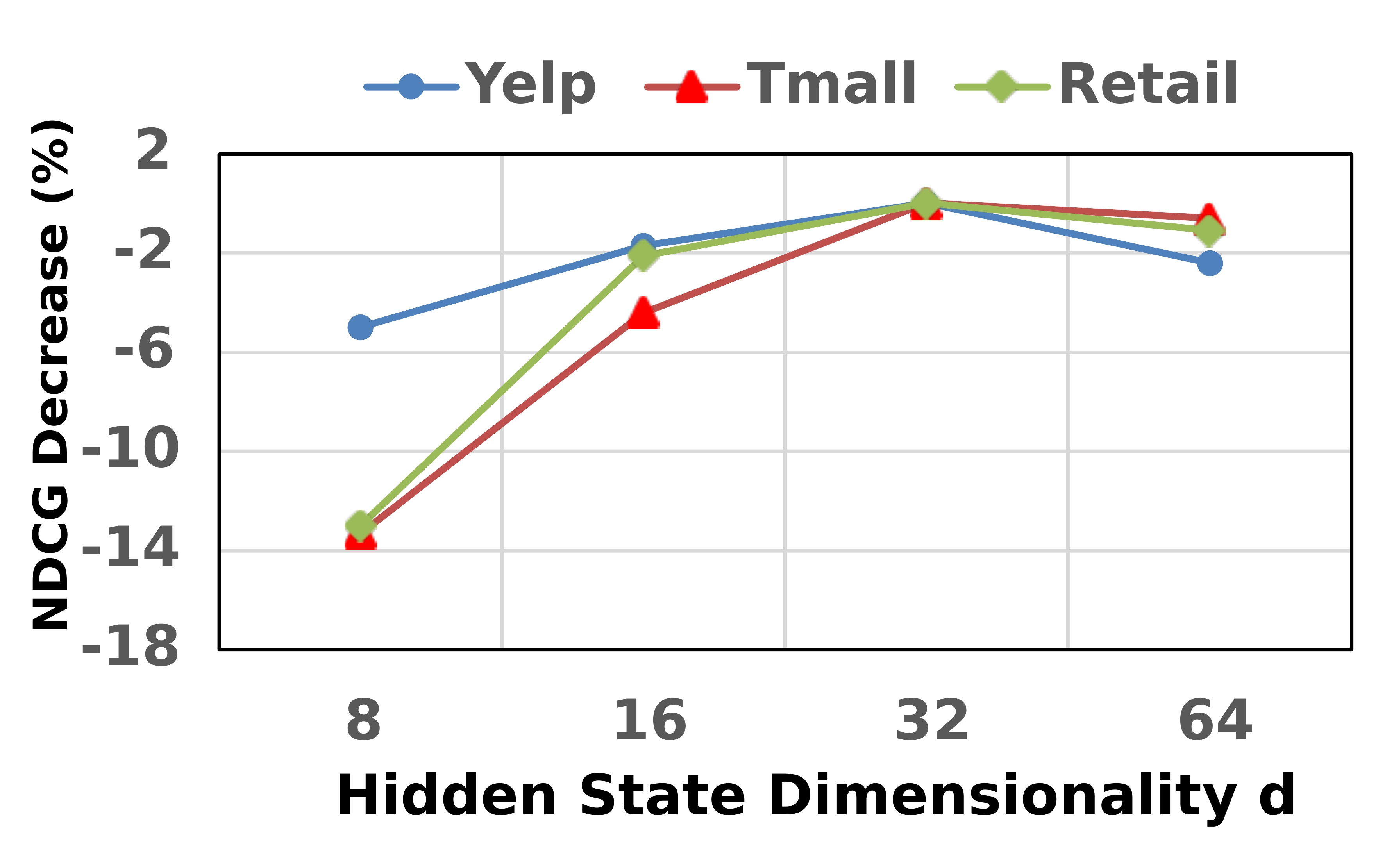}}
 \vspace{-5mm}

  \subfloat{
    \includegraphics[height=2.5cm,width=4cm]{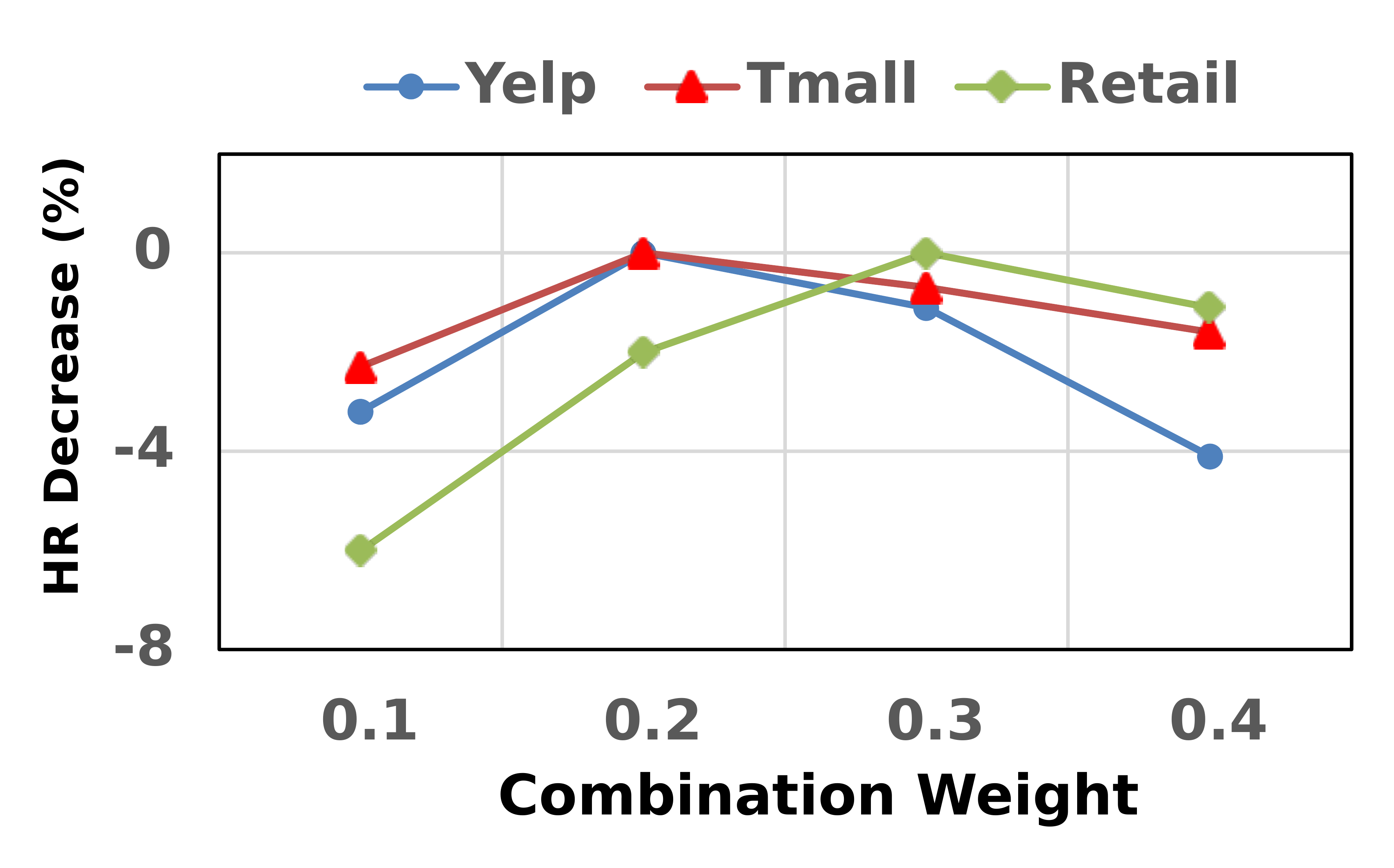}}
  \subfloat{
    \includegraphics[height=2.5cm,width=4cm]{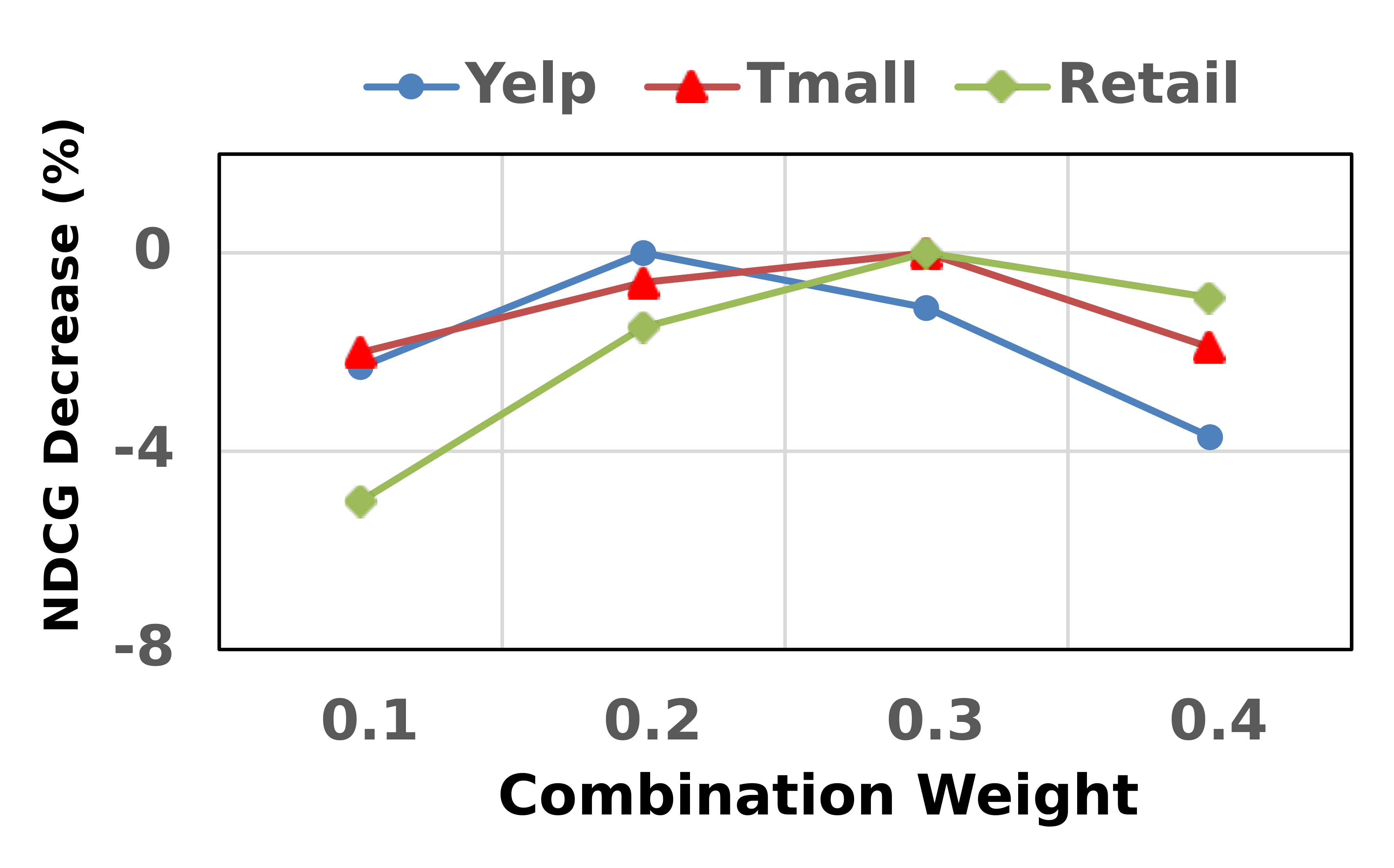}}
 \vspace{-3mm}
  
  \caption{Hyperparameter analyses of KMCLR.}
  \label{Hyperparameter fig}
  \vspace{-1.0em}
\end{figure}
\subsection{Hyperparameter Analyses (RQ4)} 
To study the effect of hyperparameter settings on KMCLR, 
we conducted experiments to explore the influence of the hidden state dimensionality $d$, graph propagation layers $L$, 
and the combination weight $\alpha$ on the results under different configurations. 
The results are shown in Figure \ref{Hyperparameter fig}, 
where the y-axis represents the degree to which the performance decreases/increases under other settings compared to the best performance with the optimal settings.
\\$\bullet$ \textbf{Hidden State Dimensionality $d$:} We vary the embedding dimensions in $\left\{8, 16, 32, 64\right\}$, and keep other optimal hyper-parameters unchanged. We can observe that the performance of the model improves with the embedding dimension increases from 8 to 32. 
However, larger embedding dimension does not always cause to stronger model representation, which leads to model overfitting.
\\$\bullet$ \textbf{Graph Propagation Layers $L$:} By stacking graph propagation layers, 
the model captures potential dependencies from higher-order neighbors, 
fully learning the collaboration relationship between higher-order nodes. 
It can be seen from the experimental results that when $L$ is 3, 
the model achieves the optimal effect. When stacking more propagation layers further, 
more noise signals may be introduced, which leads to over-smoothing \cite{42}.
\\$\bullet$ \textbf{Combination Weight $\alpha$:} To distinguish the importance of multi-behavior information and KG information, 
we set $\alpha$ in the interval $\left\{0.1, 0.2, 0.3, 0.4\right\}$. Through the experimental results, it is found that when $0.2\leq \alpha \leq 0.3$, 
the model achieves the optimal effect. 
It proves that too much KG information will add additional noise, 
while too little will lead to insufficient information enhancement.
\subsection{Performance on Alleviating Noise Effect (RQ5)} 
\begin{figure}[!htb]
  \vspace{-1.0em}
  \centering
  \subfloat[{\centering{Tmall HR@10}}]{
    \includegraphics[height=3cm,width=4cm]{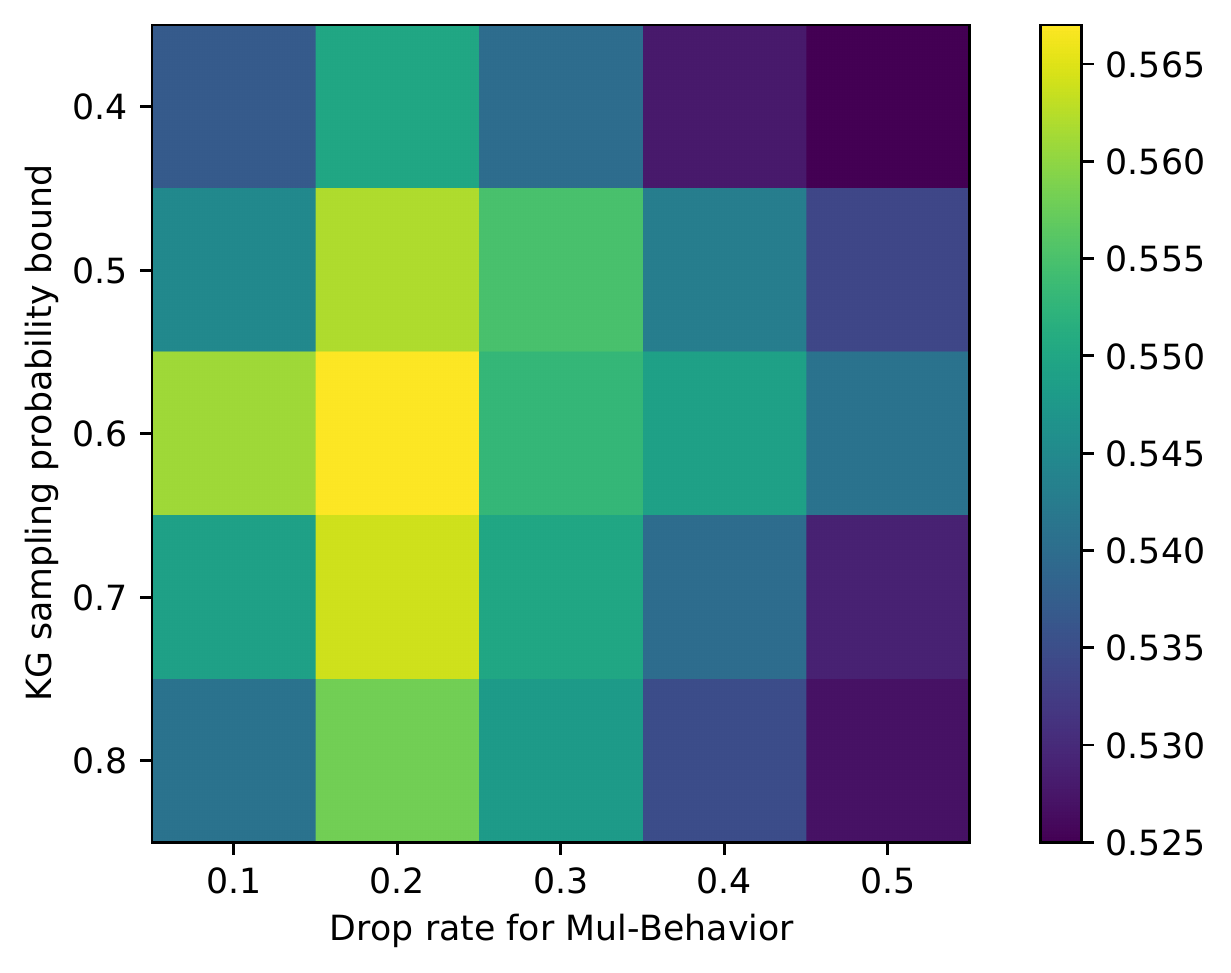}}
  \subfloat[{\centering{Yelp HR@10}}]{
    \includegraphics[height=3cm,width=4cm]{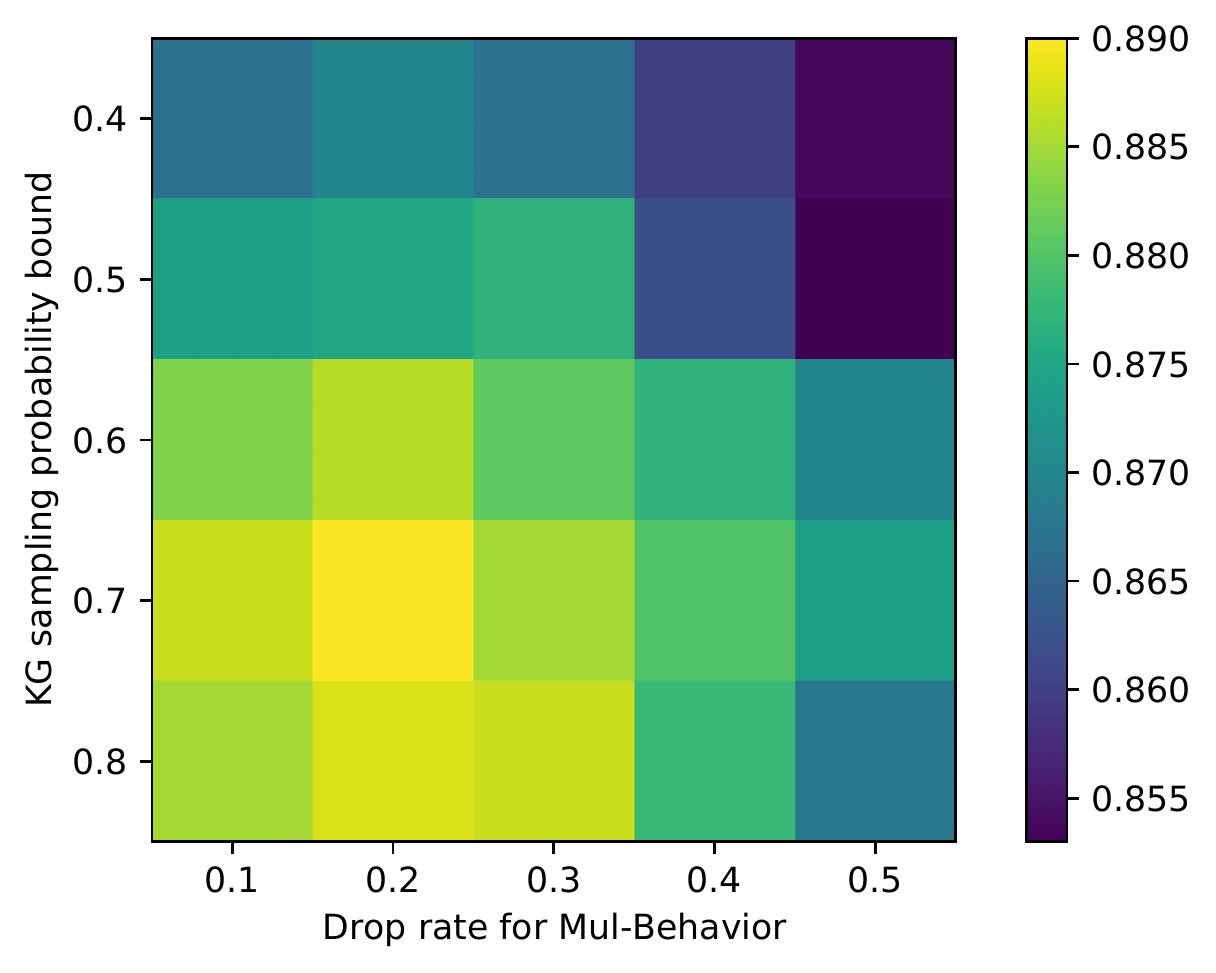}}
  \vspace{-1.0em}
  
  \caption{Parameter settings to alleviate noise.}
  \label{scatter}
  \vspace{-1.0em}
\end{figure}

While the additional auxiliary information is beneficial to improve the recommendation performance, it also leads to the introduction of noise signals. Whether in the multi-behavior interaction graph or KG, there are some invalid nodes that interfere with the final recommendation. Therefore, while using auxiliary information for information enhancement, KMCLR considers adopting methods to reduce the influence of these noise signals. Specifically, for multi-behavior interaction graphs, we set a series of random dropout probabilities to explore optimal sampling strategies under different datasets. By adjusting the interval parameters $[a, b]$ in Equation \ref{eq15}, KMCLR realizes the automatic noise reduction sampling of KG. The interval range reflects the tolerance of the model to the noise signals of KG. We conduct model noise reduction experiments on two datasets, and the experimental results are shown in Figure \ref{scatter}. The abscissa represents the random dropout rate of multi-behavior information, and the ordinate indicates the lower limit of knowledge noise tolerance. 
\section{RELATED WORK}
\subsection{Multi-behavior Recommendation}
To enable recommender systems express users' preferences more comprehensively, 
many methods propose to utilize side information of users/items to represent various relationships \cite{32}.
Therefore, recent work has begun to focus on multi-behavior recommendation, 
attempting to assist the recommendation of target behavior through individual additional behavioral information.
For example, work \cite{11} characterizes the relationship between behaviors by designing an attention mechanism. 
MBGCN \cite{13} uses graph convolutional network to learn the similarity of behaviors to discriminate individual habits.  
By combining meta-learning with multi-behavior messages, 
MB-GMN \cite{14} explores hidden embeddings behind behaviors from a low-rank perspective.
While our model excavates the inherent semantic relationship by contrasting different behaviors, 
fully considers the correlation between behavioral cognition, 
and compensates for sparse data in the form of auxiliary signals. 
\subsection{Knowledge Graph-based Recommendation}
In recent years, KG has become an important data source to enhance the existing recommendation system framework, 
thus a series of recommendation system methods based on KG have been gradually derived \cite{33,34}.
For example, the classic KGAT model \cite{29} integrates KG information and interaction graph information, 
performing recommendation prediction through graph attention network.
While KHGT \cite{25} attempts to introduce the KG information on the item side into the multi-behavior recommendation, 
and merges it into the entire interactive item representation as a single column of node embeddings.
In addition, other methods attempt to utilize KG information to enhance semantic representation, 
transforming graph information through mapping or other ways, 
which have also achieved some results \cite{35,36}.
\subsection{Contrastive Learning for Recommendation}
Contrastive learning aims to discriminate the similarities and differences between objects through pretext tasks, 
learning individual characteristics by contrasting positive and negative samples from different views. 
Contrastive learning has been demonstrated to be effective in various fields such as computer vision \cite{37}, natural language processing \cite{38}, etc. 
Recently, several works have tried to introduce self-supervised learning into recommender systems, 
and achieved some results \cite{40,41,43}.
SMIN \cite{41} proposes a heterogeneous graph neural network based on Meta-Relation, 
which is jointly trained by maximizing the mutual information between local features and global features through self-supervised learning.
The work \cite{39} utilizes self-supervised learning for sequence recommendation based on self-attention mechanism. 
It exploits the correlation of original data to construct self-supervised signals and enhances data representation through pre-training methods to improve sequence recommendation.
And our work focuses on exploring behavior semantics and mining knowledge representations, 
proposing a new CL framework that combines multi-behavior modeling and knowledge modeling.

\section{CONCLUSION}
In this work, we propose KMCLR, a novel recommendation framework for multi-behavior contrastive learning based on knowledge enhancement. 
Specifically, we jointly introduce messages of multi-behavior and knowledge graph into the recommender system to enhance the information from the user behavior side and item side, 
alleviating the issue of target-behavior data sparsity. Besides, through designing two CL tasks and a joint training paradigm, 
KMCLR can learn individual preferences from multiple perspectives, 
taking into account the commonalities and differences between different behaviors and views. 
Extensive experimental results on various datasets demonstrate the effectiveness of KMCLR compared to various state-of-the-art methods. 
A promising direction for future work is to further explore a efficient model to address the noisy signal issue caused by the introduction of extra information.

\section{ACKNOWLEDGMENTS}
This work is supported by National Natural Science Foundation of China (62003379), the“14th Five-Year Plan”Civil Aerospace Pre-research Project of China (D020101), Australian Research Council Future Fellowship (FT210100624), Discovery Project (DP190101985), Postgraduate Research \& Practice Innovation Program of NUAA (xcxjh20221605).

\bibliographystyle{ACM-Reference-Format}
\bibliography{sample-base}

\end{document}